# Conservative AI and social inequality:
# Conceptualizing alternatives to bias through social theory

Mike Zajko[1]

**Abstract**

In response to calls for greater interdisciplinary involvement from the social sciences and humanities in the development, governance, and study of artificial intelligence systems, this paper presents one sociologist's view on the problem of algorithmic bias and the reproduction of societal bias. Discussions of bias in AI cover much of the same conceptual terrain that sociologists studying inequality have long understood using more specific terms and theories. Concerns over reproducing societal bias should be informed by an understanding of the ways that inequality is continually reproduced in society – processes that AI systems are either complicit in, or can be designed to disrupt and counter. The contrast presented here is between conservative and radical approaches to AI, with conservatism referring to dominant tendencies that reproduce and strengthen the status quo, while radical approaches work to disrupt systemic forms of inequality. The limitations of conservative approaches to class, gender, and racial bias are discussed as specific examples, along with the social structures and processes that biases in these areas are linked to. Societal issues can no longer be out of scope for AI and machine learning, given the impact of these systems on human lives. This requires engagement with a growing body of critical AI scholarship that goes beyond biased data to analyze structured ways of perpetuating inequality, opening up the possibility for radical alternatives.

---

[1] Department of History and Sociology, University of British Columbia -- Okanagan Campus
Kelowna, BC, Canada mike.zajko@ubc.ca

**Introduction**

As the profound consequences of AI-related technologies have become more widely recognized, designers are increasingly expected to address the ethical or political consequences of the "tools" they are building (Hutson 2018), raising questions that have traditionally been in the domain of the social sciences and humanities. Scholars have documented the ways that automated decisions are depriving people of government benefits, discriminating on the basis or sex, skin color, age and numerous other forms of difference, choosing who is surveilled, who is imprisoned, or who is targeted for economic exploitation (O'Neil 2016; Gillespie and Seaver 2016; Eubanks 2017). Systems created with the promised of unbiased, objective judgment end up reproducing the biases and inequalities of the societies they are 'trained' on (Whittaker et al. 2018; Hoffmann 2019). As a result, 'fairness, accountability and transparency' has become a major research agenda in the field of machine learning (Lepri et al. 2018; Greene et al. 2019), and associated concerns are leading to the creation of new governance structures for AI systems (Turner 2019). There are now a growing number of AI ethics advisory panels, standards and codes of conduct (Jobin et al. 2019), and the first public policies at all levels of government – from municipal bans on facial recognition (Barber 2019) to international agreements (Simonite 2020).

Technologists are often poorly prepared for these considerations, since training in ethics has been a relatively recent complement to some computer science and data science programs (Singer 2018). If 'paradigms' are the common ways of problem-solving, or the "scientific habits" (Masterman 1970) within a discipline, then dominant paradigms in data science have been criticized as narrow technical approaches to social problems, necessitating involvement from additional perspectives (Green and Hu 2018). Social sciences and humanities scholars have been identified as having salient skills and insights to contribute to AI's holistic development (Hartley 2017; G7 Science Academies 2019), particularly in respect to ethics, fairness, and bias (Lepri et al. 2018; Silberg and Manyika 2019; Kusner and Loftus 2020). I argue that one way interdisciplinarity can benefit these discussions is by pushing beyond the conventional understanding of bias in order to better articulate the social good and the obstacles to achieving it. If politics is primarily about power, then traditional approaches in computing and data science are politically conservative in that they affirm existing power relations. AI has the potential to disrupt various institutions and social processes, but is typically used as a tool to reinforce the status quo and benefit those at the center, rather than the margins. This paper explains why conservatism is a problem and discusses conceptual alternatives to bias and equality that are better suited for radical approaches to data science.

**When systems are biased**

Today, AI is a label commonly applied to a range of technologies employing statistically-based machine learning (ML) techniques. These techniques, the data they draw upon, and the models they produce are often said to have a problem with bias. Complicating this problem is the fact that AI researchers do not agree on what bias is or what to do about it. A recent example illustrating this controversy relates to an ML system that 'upsampled' non-white faces (such as Barack Obama's) into white faces – an outcome that was then described as biased and racist. AI researchers discussing the finding on Twitter included Yann LeCun, whose initial explanation was that "ML systems are biased when data is biased" (see Johnson 2020). Similar statements have been made many times by AI developers to draw attention to the data that ML systems are trained on, and to argue that problems in the underlying data will be reproduced in the resulting system. But coming as it did from a leading figure in AI, without acknowledging the many scholars whose critiques have gone beyond database bias, the tweet triggered a contentious response. Hundreds of voices weighed in over the following days, some in support of LeCun's technical explanation of how "societal bias" is integrated into data, while others accused LeCun of ignoring and gaslighting marginalized voices, or mansplaining to experts on race and AI.

The ensuing Twitter exchanges were heightened by ongoing worldwide protests against racism and anti-Black discrimination, but as a number of participants remarked ([Lindsay 2020](#); [Raschka 2020](#)), individuals were also approaching the topic with very different implicit understandings of 'bias'. While the episode reiterated an often-repeated debate about the limitations of traditional approaches to bias in AI, the fact that similar arguments had happened many times before only served to amplify the criticism. The replies and retweets included Black AI scholars who had helped move such critiques from the field's margins and into greater visibility (such as Timnit Gebru and Ruha Benjamin). Over the previous decade, a large body of work in critical algorithm studies ([Gillespie and Seaver 2016](#)) and critical data studies ([Iliadis and Russo 2016](#)) had emerged to address the emerging structures of our digitally-mediated interactions and their relations of power, alongside a rapid development of fairness and ethics as key concerns in AI and ML research ([Jobin et al. 2019](#); [Mitchell et al. 2020](#)). And yet, many AI researchers and developers remained unfamiliar with this work, content to see 'the data' as a source of bias, and treating underlying social inequalities as out of scope.

This is precisely to sort of issue in AI research that interdisciplinary contributions from social science and humanities scholars were meant to address, and in this article I want to explore the issue of bias in AI from my perspective as a sociologist. In doing so, I provide some conceptual clarification of bias and alternative ways of understanding the problem – chiefly as one of social inequality. Inequality has a history, and a sociological analysis can help us understand the sources of un-fairness in society ([Rosanvallon 2013](#); [Machin and Stehr 2016](#)). Before moving to address bias and inequality specifically, it is important to clarify what necessitates this sort of interdisciplinary engagement, and what AI scholars can expect from interdisciplinary engagement on social issues.

**Interdisciplinarity in AI and data science**

The primary problem that interdisciplinary approaches to data science and AI development can address is one of scope, to compensate for a traditionally narrow, formal approach to problem-solving in AI ([Green and Hu 2018](#)). Explanations and critiques from social sciences provide a 'helicopter view' by rising above a particular situation to get a wider perspective, then carrying these insights down to deal with on-the-ground specifics ([Thompson 2017](#)). Social science theories operate at different levels of analysis, and can ideally connect local circumstances with general, structural, macro-level phenomena. This means situating AI within larger historical and social forces, including contemporary forms of capitalism ([Srnicek 2017](#); [Penn 2018](#)), white supremacy ([Benjamin 2019](#)), settler colonialism ([Mohamed et al. forthcoming](#); [Kukutai and Taylor 2016](#)), gender binaries and heteropatriarchy ([Keyes 2018](#); [Costanza-Chock 2020](#)), as well as governing rationalities ([Rouvroy and Berns 2013](#)), international relations, ideologies and cultural representations. The push to develop and commercialize technologies occurs within a specific political economy ([Kushida 2015](#); [Hoffman 2017](#); [Crawford and Joler 2018](#); [Cohen 2018](#)). A business model's legitimacy depends on cultural attitudes and how capital is organized ([Levina and Hasinoff 2017](#)). Internationally, quasi-sovereign platforms and nation states either court one another, or fight for autonomy ([MacKinnon 2012](#)). AI introduces some new elements into these dynamics, but technologies are the product of existing social arrangements, and their success or failure depends on far more than their ability to solve problems, or whether they meet the needs of users.

While there is extensive literature in the social sciences on many of these topics, due to the ways that the disciplinary "divide" is reinforced, these works have been easy for data scientists and AI developers to ignore (Moats and Seaver 2019). In addition to the fact that they often write for different audiences, it is also important to note that data scientists and social scientists are often trying to answer very different kinds of questions through their own methods (Wallach 2018). This leads to challenges in translating literature from one domain to the other, and different ideas about what an interdisciplinary approach would look like.

In *multi*disciplinary work, knowledge is added or applied from one discipline to another (either to add a dimension or as 'context'), but disciplinary boundaries can still be maintained and reproduced (Klein 2017). In contrast, scholars of *inter*disciplinarity tend to emphasize more of a reciprocal, transformative, or integrative relationship between disciplines. According to Lury, interdisciplinary methods are "dynamic conduits for relations of interference in which differences and asymmetries between disciplines are explored and exploited in relation to specific problems, in specific places, with specific materials" (2018, p. 21). While in Lury's formulation, interdisciplinary work is oriented towards specific problems (particularly "global problems"), the emphasis is on crossing disciplines to 'constitute' problems in new ways, rather than arriving at novel solutions. In this conceptualization, interdisciplinarity is less an instrumental practice and more of an enterprise in research autonomy (Lury 2018). In contrast, other writers have identified varieties of instrumental interdisciplinarity that do indeed serve the needs of a specific field, commercial imperatives, or national goals. As Klein (2017) notes, these approaches may include the possibility of critique, but there is a tension between autonomous, critical forms of interdisciplinarity that take a broad view, with instrumental approaches to solving economic and technical problems.

Part of the tension is that dominant approaches in data science avoid being overtly political, and in the attempt to remain neutral, objective, and fair, end up taking an implicitly conservative approach. As Green and Viljoen (2020, p. 22) argue, "This emphasis on objectivity and neutrality leads to algorithmic interventions that reproduce existing social conditions and policies. For objectivity and neutrality do not mean value-free—they instead mean acquiescence to dominant scientific, social, and political values". Efforts to promote ethics in AI development are therefore criticized as "ethics-washing", "empty gestures" and "weasel words" (Metcalf et al. 2019), while explicitly political efforts to challenge social structures remain on the margins of the field. Conservatism remains the default position, while radical alternatives require justification.

**Conservatism and radicalism in AI**

Rather than seeing conservatism as a psychological trait, political identity, or set of policy positions, I use the term to refer to the conservation and maintenance of the social and political and economic order. Conservative approaches to AI "optimize the status quo" (Carr 2014), promoting changes that preserve existing inequalities. Abeba Birhane (2020) has been one of very few to actually describe this tendency as 'conservative', but there is widespread recognition that AI, as it currently exists, reproduces existing social distinctions and biases. The fact that biases in data end up being replicated in models has been used to critique AI, but claims about biased data can also be used to deflect responsibility and defend the enterprise, by situating the source of bias as being outside of AI and thereby out of scope (see Hoffmann 2019). Meanwhile, various radical approaches are actively being developed in opposition to these conservative tendencies (Cifor et al. 2019; Gurumurthy and Chami 2019; Keyes 2019; Doyle-Burke and Smith 2020; Costanza-Chock 2020; Kalluri 2020), but conservatism remains entrenched in AI through longstanding proximity to powerful interests.

The early history of Silicon Valley includes radical visionaries and countercultural ideas, and the companies that emerged to disrupt entire industries in the 1980s and 90s often promoted radical change as a core value. However, this was an individualistic, libertarian form of radicalism, which was "more apolitical than ideological" (Johnson 2018) in the sense that it avoided formal politics and pursued world-changing through private initiative. In other words, the radical tendencies of Silicon Valley proved compatible with capitalism (Streeter 2011), creating an "awkward fusion of market values and vague humanitarianism" (Silverman 2018). By the mid-2010s, when dot-com survivors and new startups spoke about being agents of radical change, their claims were ridiculed on HBO's satirical series *Silicon Valley*, or the writings of Evgeny Morozov (2013). People became increasingly skeptical of the claims that new technologies would radically improve the world, and ethical failures among leading technology companies were widely reported. Today's giants of Silicon Valley are major centers of power in society, and the existing order serves them just fine. In this context, an inherently conservative approach is entirely predictable in its emergence and its consequences; virtual assistants like Alexa are not created to assist people in challenging capitalism or state power, and AI companies pursue major government and corporate contracts.

Today's new AI startups are "are acquisition targets not only for big tech companies, but also for traditional insurance, retail, and healthcare incumbents" (CB Insights 2018). There is no shortage of radical ideas, but the ones that receive lasting support end up being those that reinforce dominant interests. As Ted Chiang (2017) and Jonnie Penn have argued, AI today follows the logic of capitalism and "thinks like a corporation" (Penn 2018), which determines the priorities for AI development. Access to resources are crucial; The non-profit OpenAI, in its quest to develop artificial general intelligence (AGI) to change the world, eventually decided to find a wealthy corporate investor (Microsoft) so that it could exchange "capital for computational power" (OpenAI 2019).

Academic research can be more autonomous from government and corporate priorities, tolerating more radical research agendas, but institutionalized cultures in academia have also often reinforced a narrow focus (for instance, through the formalism in computer science, see Leith 1990; Selbst et al. 2019; Green and Viljoen 2020), and academic projects must be carefully aligned with sources of funding, often including close ties with private industry (Hoffman 2017). Finally, computer science and data science have long been recognized as having a 'diversity problem', specifically through limited participation by women, other gender minorities, and Black scholars. This results in a "feedback loop" through the design of AI systems that produce unequal and discriminatory effects (West et al. 2019).

**The consequences of AI conservatism**

As I hope the above makes clear, when I discuss conservatism I am not referring to the various values that have historically been associated with conservative thinkers and political movements. People and organizations that self-identify as conservative do sometimes champion change, and political theory is ill-served by dividing ideas and movements into univocal, opposing camps (Bourke 2018). Understanding actual political struggles in their historical context means dispelling with the notion that political actors can be categorized or essentialized according to coherent, contradictory positions, such as conservative and radical (Edwards 2006). People who self-identify as conservatives have often held revolutionary sentiments, and political radicals are not agents of chaos; "even anarchism aims to maintain its preferred values, if not the state as a vehicle to secure them" (Bourke 2018, p. 453).

Keeping in mind that this is not representative of the complexity and contradictions of conservatism and radicalism as these exist in actual political struggles, I use conservatism and radicalism in a simplified heuristic sense to distinguish between conserving the existing order or radically changing it to something new (see Wolfe 1923). While my goal is to use these terms for their analytical value in explaining the political stakes of AI in society, this can be difficult given the moral values frequently attached to conservatism and radicalism. Furthermore, I am making a normative argument of my own, in that I believe an implicitly conservative orientation is harmful as the default position in design. This is not because I think conservatism is inherently bad or harmful; there are aspects of the existing order that are worth keeping or reinforcing, but there are also many things about society that should be changed or improved. The problem lies in the limitations that a conservative approach imposes on social change and the possibilities it forecloses, what an economist would consider the 'opportunity cost' of the road not taken. This results in "algorithmic interventions that entrench existing social conditions, narrow the range of possible reforms, and impose algorithmic logics at the expense of others" (Green and Viljoen 2020, p. 20). Radicalism on the other hand, is also neither good nor bad; radical change can be beneficial, dangerous, or necessary as a political orientation. Preserving the existing social order precludes these possibilities.

As a concrete example, I am writing this following widespread calls in the U.S. to defund or disband police forces, amid scenes of police violence against peaceful protests, sparked by the killing of George Floyd by police officers in Minneapolis (Pettit 2020). While police abolition draws on a long intellectual history (McDowell and Fernandez 2018), it remained a radical aspiration until an accumulation of visible injustice made structural changes to existing systems a real possibility in 2020. Work on algorithms for justice and security (Hayes et al. 2020) has often been critical of their misuse by state authorities, but focused on making "reformist" improvements to the existing system (Green 2019), rather than questioning the centrality of the system for upholding social order. Such fundamental critiques and abolitionist arguments have been made from outside data science, or within critical data studies (Benjamin 2019; Green 2019; Costanza-Chock 2020), but conservative currents in AI and data science remain strong, favoring better data as a solution to injustice.

Political alternatives to state-centric ideas of justice and security are inconceivable as long as we rely on some fuzzy idea of the social good that predominates in discussions of data science and AI. Instead of considering what an ideal world of justice and security would look like, AI is often tasked with addressing "rough proxies", such as reducing crime, thereby "accepting and working within the parameters of existing systems to promote the achievement of their goals" (Green 2019, p. 3). If justice and security are served by creating more accurate and efficient tools for law enforcement, what happens when agents of the state create insecurity and injustice? Among AI developers and their critics, this is a topic of intense controversy. The potential for algorithms and data sets developed at Google, Amazon and Microsoft to be used for oppression and injustice have led to critiques and resistance from employees (Campbell 2018). But how do we go beyond recognizing the harms caused by algorithm-enabled state violence and move toward an alternative? If we can recognize the unethical use of AI when we see it, what does the opposite look like? In 'ethical AI', the answer has typically been presented in terms of increasing fairness and reducing bias. Examining the weaknesses of this approach will also allow us to consider some alternatives.

**Debiasing society**

In what follows, I offer an interdisciplinary contribution to the topic of fairness and bias in AI, where interdisciplinarity is used to formulate new problems, rather than providing new solutions to existing problems. Even in situations where two disciplines use different language to discuss what might appear to be the same problem, switching from one disciplinary discourse to another can significantly shift how problems are formulated. This is certainly the case when it comes to questions of bias and fairness in AI, once we attempt to translate these into discourses on inequality in the social sciences and critical theory (Sloane 2019). For instance, 'reducing bias' is not at all synonymous with 'reducing oppression'. Researchers can attempt to correct for "societal bias" (Friedman and Nissenbaum 1996; Silberg and Manyika 2019), but understanding the source of this structural inequality is typically treated as irrelevant, or outside the scope of analysis. Social structures, or "sources of discrimination that cannot be traced to discrete bad mechanisms are bracketed, dismissed as someone else's problem or, worse, couched as untouchable facts of history" (Hoffmann 2019, p. 905). As a result, the discourse of fairness in AI is far removed from the sorts of measures being discussed to oppose systemic racism, gender-based oppression, or colonialism.

On the one hand, this is part of a recurring critique of AI development which argues that narrow technical approaches "miss the broader context" (Selbst et al. 2019, p. 59). When social concepts are incorporated into data science, social issues can become a mathematical abstraction, as when researchers "reify fairness as social concept into fairness as satisfiable technical criterion" (Green and Hu 2018, p. 1). The problem here is that "to treat fairness and justice as terms that have meaningful application to technology separate from a social context is… a category error" (Selbst et al. 2019, p. 59). Selbst et al.'s STS-informed analysis provides valuable "remediation strategies" for five "failure modes" in fairness research, broadly through deeper engagement with "the social" to understand ML systems as sociotechnical (Selbst et al. 2019, p. 59). However, the solution is not simply to add a missing layer of social context. Particularly if we go beyond STS scholarship on sociotechnical systems and consider more general sociological theories of inequality, stratification, and power (Grusky 2018), deeper engagement with 'the social' means displacing or superseding the conventional meaning of bias. To consider what this would mean, it is important to review the concept of bias in data science, ML, and AI ethics.

**Conceptualizing bias**

The initial challenge in understanding how bias is currently conceptualized is that the term has multiple potential meanings, and many authors do not bother to define it. Rather than referring to a commonly-accepted definition of bias, the best approach is to consider how the term is commonly used, or what 'work' the concept is typically doing. In literature on data science, machine learning and AI, the implicit general definition of bias is any tendency, pattern or association that is problematic. There are examples of texts that discuss "useful bias", or biases that are desirable, particularly when discussing some human predispositions (Shah et al. 2019b), but generally when bias is invoked, it is a problem.

Some texts do invoke specific forms of bias, which are typically differentiated by their source: human bias, machine bias, systemic bias, societal bias, historical bias, sampling bias, observation bias, and so on (for example, Shah et al. 2019a). There is considerable conceptual confusion and overlap with these terms, but they typically distinguish where the bias is supposedly 'coming from'; whether an individual decision maker's unconscious bias, historically entrenched distributions, or the issues with data collection and measurement. Whatever the source, all relate to at least one fundamental definition of bias, as being either (1) inaccurate, or (2) undesirable. This aligns with the distinction made by Mitchell et al. (2020, p. 4) between "*statistical bias*—i.e. concerns about non-representative sampling and measurement error—and *societal bias*—i.e. concerns about objectionable social structures that are represented in the data".

The most straightforward conceptions of bias to operationalize are statistical in nature, and based on the accuracy of representations or predictions – i.e., does the data accurately reflect the 'ground truth'? Does the output of the algorithm accurately reflect the risk profiles or recidivism rates for different populations? Is the person with the best qualifications for the position accurately identified, or does some irrelevant factor bias the decision against them? These are questions that can be addressed with better data and models, overcoming problems with sampling, measurement, and design. However, the vast amorphous terrain of societal bias includes inequalities and injustices that can indeed be accurately reproduced, and therefore reinforced, by an algorithm. These include cases where the "training data is tainted with historical bias" or where there is an "unequal ground truth" ([Hacker 2018, p. 1148](#)).

Gender bias exists even where an algorithm accurately reproduces cultural assumptions about gender embedded in language, returns photos of men for an image search of an occupation that is male-dominated, or predicts that an individual is a man on the basis of their occupation (see [Suresh and Guttag 2020](#)). It is sometimes possible to reframe these issues in terms of accuracy – perhaps the algorithm does not accurately reflect some fundamental equality between people or groups – but when authors are concerned about algorithms that 'reproduce bias', this can include the accurate reproduction of existing social patterns and distributions. In these conceptions, bias is taken to mean whatever is "unfair", "undesirable", or "unwanted", which typically manifests as "systematic discrimination… based on the inappropriate use of certain traits or characteristics" ([Silberg and Manyika 2019, p. 2](#)). The easiest way to operationalize this second definition of bias is to tie it to some conception of fairness – hence any tendency we can consider unfair is also biased ([Friedman and Nissenbaum 1996, p. 343](#)). However, this does not get us very far because of multiple contradictory conceptions of fairness ([Friedler et al. 2016](#)).

Defining bias in terms of (1) accuracy of prediction/estimation is well-suited for technical problem-solving, but not for addressing the widespread concern over 'reproducing biases' in society ([Hoffmann 2019](#)). The most inclusive (2) definition of bias is a tendency that is undesirable or unwanted, based on a "normative concern with the state of the world" (Suresh and Guttag 2020, p. 2), but this tells us nothing about what is desirable. Addressing this kind of bias involves either selecting one of the formal definitions of fairness, or some ad-hoc notion of a desirable outcome. While 'minimizing bias' can produce positive or progressive outcomes, this is not equivalent to social justice. The language of bias does not help us attend to how capitalism, patriarchy, white supremacy, or colonialism are organized, and actually makes this work more difficult. Bias places the focus on circumstances of disadvantage, rather than "the normative conditions that produce – and promote the qualities or interests of – advantaged subjects" ([Hoffmann 2019, p. 907](#)). In other words, we end up focusing on systematically disadvantaged groups, rather than the reasons why these disadvantages exist (to produce systematic advantages for others). These are the situations in which social theory cannot be bracketed, and where the language of bias needs to give way. Interdisciplinary engagement cannot flesh out the missing details of 'societal bias', but it is helpful to problematize the concept.

The argument I am making here is not just to critique conceptions of bias in AI ethics, but to show the challenge of interdisciplinarity on a specific topic where insights from social science and humanities scholars are presumed to be valuable ([Kusner and Loftus 2020](#)). And indeed, these outsider perspectives do provide value – just not in the instrumental way favored for multidisciplinary collaboration in data science. Instead, this kind of conflict is entirely consistent with an understanding of interdisciplinarity as "interferences between disciplines", rather than some harmonious synthesis ([Lury 2018, p. 1](#)). A critical interdisciplinarity "interrogates" knowledge "with the aim of transforming it" ([Klein 2017, p. 28](#)). Societal bias ([Friedman and Nissenbaum 1996](#)) and related concepts like human bias, historical bias or 'unequal ground truth', could benefit from being transformed or replaced by more elaborated concepts in social theory.

Social science literature can help with conceptual work around bias (1) as accuracy, including practical lessons that research methods can offer for operationalizing the sorts of concepts social scientists have long been measuring ([Goertz 2006](#); [Babbie 2014](#)). STS and the philosophy of science provide some critical perspectives on ontology, representation, and objectivity ([van Fraassen 2008](#); [Jasanoff 2011](#)). However, the greatest need for new approaches is in definitions of bias (2) that are based on some fuzzy notion of "world as it should and could be" ([Mitchell et al. 2020, p. 4](#)), so that we can articulate just what we find undesirable and what we want to produce in its stead. Rather than treating bias as something that can removed from the world, we need to more directly consider what kind of world we want to create, and to this end we are better off relying on conceptualizations of inequality, structured distributions, and power.

**Working towards equality**

The issue of structured inequality should force us to return to the second definition (2) of bias discussed above. Specifically, we need to consider what kinds of inequality are undesirable, and to ask the politically contentious question of what kind of outcome we actually want. Is it a society with a more equitable distribution of resources and wealth? Alternately, is it a society where individuals have greater equality in their opportunities to obtain these unequal rewards? The second goal, equality of opportunity (EoO), can be defined in simple terms as the view that individuals should have an equal opportunity to achieve success, rewards, or to obtain resources ([Ferreira and Peragine 2015](#)). This does not result in equal outcomes, but in inequalities that are just or fair if they are derived from individual efforts or some related notion of 'merit'. In this view, it is unfair, biased, or discriminatory to structure opportunities based on circumstances that are beyond the individual's control.

EoO has been cited as a justifying rationale to "remove discrimination" ([Hardt et al. 2016](#)), but there are limits to the bias and discrimination that can be subtracted when dealing with structural forms of inequality. Much depends on what we consider to be relevant when determining whether two individuals are equal, or where we assign responsibility for unequal outcomes. It is one thing to ensure that candidates with equal qualifications have an equal opportunity to be successful, and another to address why some kinds of people tend to be more qualified than others. A formal approach to EoO will conserve many structural inequalities that exist prior to the opportunity being considered.

EoO also necessitates a choice about when individuals (specifically, children) are equal and autonomous enough to be responsible for the outcomes of their choices. According to Roemer, determining what aspects of an individual's behavior are considered within or beyond their control is a question for each society to decide "through some political process" ([1998, p. 8](#)). For example, it is often considered fair to make hiring decisions based on a candidate's achievements in school, but unfair to evaluate merit on the basis of other factors outside their control. No one chooses the circumstances into which they are born, but the quality of schooling that a child receives can depend on where they live, as well as race, gender, and disability. Disadvantages early in life can have cumulative effects later on, while advantages can be used to beget further advantages ([Rigney 2010](#)). Sociologists of education and class (among whom Pierre Bourdieu is best-known) have shown how class inequalities are persistently reproduced even in educational systems established to dissolve class boundaries. Educational outcomes depend largely on family resources, social class, and gender-specific familial expectations ([Archer et al. 2003](#); [Yamamoto and Holloway 2010](#)). There is no point in life where these determinants become irrelevant and a person becomes fully autonomous and independent of factors beyond their control, whether these are historical circumstances, luck or 'pure chance' ([Gandy 2009](#); [Segall 2013](#)). But a hiring or sentencing algorithm cannot disentangle personal agency from social structure, so instead we have systems that rely on 'protected categories' of personal attributes, arbitrary delineations of when adulthood begins, and a presumption of personal responsibility that aligns with political, legal, and economic norms.

There are different ways of designing algorithms to maximize equality given these problems, but any assumptions made in operationalizing fairness have political implications. This is because "algorithmic fairness… reproduces, at a technical level, the tensions inherent in the political philosophy of justice" ([Hacker 2018, p. 1183](#)), such as the tension between individual and group-level fairness, anticlassification and antisubordination theory ([Barocas and Selbst 2016](#)), or the related tension between EoO and substantive equality (SE). Where EoO tries to level the playing field between individuals, approaches based on substantive equality (SE) are explicitly intended to overcome the conservative limitations of EoO through a deeper view of inequality and the use of positive action. This means appreciating that for members of "socially disadvantaged groups, the game of taking part in social participation is probably rigged" from birth (Baumann and Rumberger 2018, p. 5). If group members experience inferior access to education, and if equal opportunity is based on equal education, then educational resources for disadvantaged groups must be improved. Where 'right-wing' conservative interpretations of EoO tend to "limit the scope of circumstances" being considered and "attribute much to individual choice", arguments on the political left emphasize how social circumstances determine 'choice' (Ferreira and Peragine 2015, p. 5). SE necessitates redistributive policies, going beyond bias and discrimination as factors to be removed, and affecting those who currently benefit from social inequality. EoO has more conservative implications, preserving many structural inequalities.

Equality then, is neither a singular concept, nor a moral or philosophical given. While equality does depend upon some sort of mutual recognition of commonality, the nature of this shared character (the way in which we are supposedly equal), and the idea of a social order based on equality, both have a history. In other words, equality is an idea that had to be invented (Rosanvallon 2013), and which has been invented in different ways. Political philosophy can help us think through the problems of equality and fairness (Binns 2018), and provides no shortage of divergent answers. But history is instructive to show us that we are not operating in some abstract political and theoretical space. Not only can we learn from the struggles of earlier generations with these same problems, but the history of ideas about equality shapes the form that these debates take today.

These are important questions to consider once we go beyond bias and fairness as technical criteria, appreciating scholarship on how inequalities are perpetuated as well as the political stakes in debates over equality. But equality also has limits as an ideal, where one kind of equality can cover up a deeper injustice, or where desirable social arrangements are best characterized in other terms. If we want to take inequality seriously, at some point we are going to have to move beyond negative ideas of removing bias, discrimination, or inequality, and to think in positive terms about what we want to create instead. To this end, it is valuable to consider some specific forms of inequality in relation to the problem of bias in AI. The following examples relate class, gender and race, which social theory now generally treats as intersecting categories that are continually produced and reproduced, including through discrimination by AI systems.

*Class and economic inequality*

There are many lines of inequality in society, and my discussion is limited to a small number for the purposes of illustration. However, the term 'social inequality' is most often used to refer to economic inequality, either between social classes, or some conceptual alternative relating to wealth and socio-economic status (Grusky 2018). The literature on bias in AI has relatively little to say about economic inequality, other than as an outcome of algorithmic decision-making. Class-based discrimination is generally treated as less problematic than discrimination according to protected categories such as race, gender, and disability (Costanza-Chock 2020). Unlike gender and racial bias, 'class bias' does not provide much of a conceptual foothold. Bias (1) can certainly be discussed in terms of inaccuracy or error, such as when some people are wrongly excluded or denied credit because of errors in their credit reports, but when people are included for targeted discrimination or exploitation, these processes tend to be discussed as "predatory" rather than biased (see Johnson et al. 2019). Conceptualizing these processes as predation or bias calls out the most exploitative dimension of capitalism, but this is not a foundation on which to proceed to a better society. A bias-free world could be one where every person is equally surveilled and controlled, or equally targeted by predatory lenders.

While we could characterize algorithms that target the vulnerable and economically precarious as being biased against the poor, it is more accurate to say that the poor are managed and controlled, based on longstanding assumptions that it is desirable to discriminate against the poor (Eubanks 2017), or that the poor are exploited by algorithms designed to "target and fleece the population most in need" (O'Neil 2016, p. 81). We therefore need to examine the structures through which the poor are kept poor, so that we can eliminate or circumvent these where possible, and strengthen the sorts of structures that could promote social mobility and economic betterment. Class, like other dimensions of inequality discussed below, is reproduced on an ongoing basis (Grusky 2018), preserving the privileges of wealth as well as the disadvantages of poverty from one generation to the next. By the time AI systems enter the picture, this unequal ground truth is well-established, intersecting with other dimensions of inequality like gender, race, and disability (Hoffmann 2019).

Building a fairer decision-making algorithm will only have a superficial impact on fundamental inequalities, but a structural analysis can identify where more systemic changes would be effective, including where automated decision-making should be removed from a process (Eubanks 2017). A radical approach goes further in reducing unequal outcomes and unequal distributions of resources, goods, or other kinds of "holdings" (Segall 2013). This also broadens the focus from disadvantages or obstacles, to considering the advantages or privileges maintained by existing structures. Instead of imagining that we can lift the marginalized up to become full participants in some 'level playing field', radical arguments often seek to overturn the competitive playing field altogether to pursue other forms of social relations.

*Beyond equality: Gender*

Compared to issues related to social class, gender bias is more familiar territory for discussions of fairness in AI. Gender equality might seem to be a more acceptable goal than equalizing wealth and resources, but here it also helps to be more specific about what we are trying to achieve. Are we trying to produce equally-accurate estimates about men and women, or cis and trans people? Are we aiming for equal gender representations or equal opportunity in currently unequal distributions? Debates in feminist theory have long included critiques of formal equality between men and women, given the ways that a liberal ideal can actually preserve structures of inequality (Bryson 2016). For example, providing men and women equal opportunities for paid employment makes little difference if women are still expected to do the majority of unpaid work. More radical perspectives link gendered inequalities with our ideas about gender and society, arguing that it is on this conceptual level where interventions should be targeted.

Rather than seeing gender as an attribute of persons, sociological theory discusses gender as a process, a 'doing', an identity, and a hierarchical system that intersects with other forms of inequality (Risman et al. 2018). An early distinction between sex and gender provided an important foundation for gender theories, but this is not as simple as basing sex (male/female) on biological categories in contrast to gender (masculine/feminine) as a fluid social construction. Sex and gender classification are both social processes, and neither is straightforward even in the absence of AI. Human physiology does not neatly differentiate people into male and female – this is work carried out within sociotechnical systems such as medical institutions, where doctors have to confront the fact that many human bodies are not easily classified in a binary framework. Historically, this has led to surgical interventions that harmfully and irreversibly attempted to eliminate ambiguities from human bodies (Davis and Evans 2018). AI systems make similar sex and gender-based classifications (often conflating the two), based on the idea that the human population can be divided into discrete, natural groups that can be 'read' by others on the basis of appearance. For example, a frequent assumption in image classification is that gender can be accurately or inaccurately classified according to a male/female binary (Scheuerman et al. 2019). Work on gender bias and fairness can address the accuracy of such classifications or whether men and women appear in equal proportions, but feminist technology studies show how technology and gender are co-produced (Wajcman 2007), and how gender is "inscribed" through technologies, including AI systems (Adam 1998).

To go beyond addressing how AI systems reproduce gender biases found in training data, we can examine how AI systems reproduce gender as a social structure – one which we might specifically identify as heteropatriarchy; through the ontological assumption that a gender binary exists, in which superior values are associated with men over women, and where heterosexuality is presumed along with traditional (cis) gender presentation. This offers many opportunities for changes that go beyond equal treatment, or equal outcomes for men and women, to examine what ideas about gender are affirmed or excluded. Radical alternatives can include destabilizing cis-normative and binary-normative structures ([Costanza-Chock 2020](#)), or 'queering' normative gender distinctions ([Cockayne and Richardson 2017](#)), and there are a number of practical ways forward toward "non-binary utopias" ([Spiel et al. 2019](#)). These are political courses of action to be sure, but affirming heteropatriarchy through the reproduction of traditional gender norms in ML systems is also political ([Costanza-Chock 2020](#)), albeit conservatively so. Feminist and queer theories are helpful in articulating desirable outcomes that go beyond gender equality as an ideal, but similar issues arise with the limitations of 'debiasing' as an approach to racism.

*Race and colonialism*

As with gender, racial categories are the outcomes of historically and culturally-contingent classification schemes that associate hierarchical values with supposedly natural groups of people ([Benthall and Haynes 2019](#)). Racial bias is one of the most prominent forms of bias discussed in AI and ML, with the example of the COMPAS recidivism prediction algorithm having achieved a paradigmatic status in discussions of the topic ([Wong 2020](#)). This literature is typically U.S.-centric, focusing on differences between Black and White Americans, and U.S.-based scholarship also provides widely influential concepts and theories for understanding race and racism. There are significant theoretical affinities as well as differences among scholars of Black radical thought, critical race theorists, neo-Marxists, and sociologists of race, regarding concepts such as white supremacy, racialization, and the relevance of class ([Robinson 2000](#); [Omi and Winant 2014](#); [Bonilla-Silva 2015](#); [Walton 2020](#)). Despite their disagreements, these approaches share a view of race and racism grounded in social structure – explanations that go beyond individual human biases. This kind of background is crucial for understanding how the 'tech-to-prison pipeline' functions in an inherently racist manner, even in the absence racist intent, and why certain forms of AI can only strengthen these structures and their oppressive effects ([Coalition for Critical Technology 2020](#)).

Although many of these arguments are made in the specific context of anti-Black racism in the U.S., they do have more general applicability to how racialized people are governed through discriminatory algorithms. Theories of colonialism are an underdeveloped means of understanding issues in AI through historical inequalities and asymmetrical power relations ([Mohamed et al. forthcoming](#)), and moving beyond approaches to AI ethics grounded in formalizing fairness and equality. The fight for equality has been historically important in anti-racist and anti-colonial political struggles, but so has the radical goal of dismantling systems of oppression and the creation of new sources of solidarity. To illustrate this, it is important to consider what it would mean to connect concerns over racial bias in AI with the well documented inequalities between indigenous and non-indigenous people, using theories of settler colonialism ([Veracini 2010](#)) and my home country of Canada as an example.

If we see our challenge as one of minimizing bias, we might look to reduce group differences across the many dimensions where there are stark inequalities between indigenous and non-indigenous populations in Canada, the U.S., and comparable nations ([Cooke et al. 2007;](#) [Cunneen 2014](#)). Certainly, there is evidence of discrimination against indigenous peoples in different societies ([Blackstock 2017](#)), but it would be mistaken to assume that this inequality can be addressed by reducing bias among those making decisions that concern indigenous lives. The common cause behind these inequalities (and the common thread in many societies in the Americas, as well as Australia and New Zealand) is not individual, societal or historical bias, but settler colonialism. This is now a well-theorized social structure, organized around dispossession, indigenous elimination, and control over land ([Veracini 2010](#)). Putting a name to the structures of inequality moves us closer to understanding the otherwise amorphous social context, why unfairness and injustice are distributed as they are, and towards normative approaches that go beyond equality.

An algorithm can address the kind of personal bias that denies an applicant a job because of their ethnicity, but this kind of discrimination is one relatively minor part of what perpetuates colonial inequalities. The 'unequal ground truth' in Canadian society for example, is one in which indigenous populations have often been governed through different institutions, receiving inferior educational opportunities, inferior access to health care and basic infrastructure, and where indigenous children are much more likely to be raised in poverty and foster care, as consequences of generations of colonization ([Truth and Reconciliation Commission of Canada 2015](#); [Blackstock 2017](#); [Statistics Canada 2019](#)). In a sociologically-informed analysis of how racial categories are reproduced in the U.S., Benthall and Haynes ([2019](#)) argue for disruptive alternatives to existing hierarchies that work towards social integration and similar treatment for segregated groups. However, social integration and universalism cannot be assumed as goals when addressing racial and ethnic inequality, especially in a colonial context. Proposing formal equality as a solution to these injustices is problematic, since treating everyone the same (for instance, as Canadian citizens with equal rights and opportunities) amounts to a form of assimilation. Rather than equality, Canada's officially-recognized objective is 'reconciliation' ([Truth and Reconciliation Commission of Canada 2015](#)), and more radical scholarship asks us to consider ways of working towards decolonization and indigenous resurgence ([Simpson 2016](#)).

Once again, by naming desired goals, or articulating a positive value to work towards, we can be more specific than the negative process of removing bias. In doing so, it is important that our normative goals are attuned to social context rather than universal notions of equality. Both decolonization and reconciliation involve recognizing the distinctiveness of indigenous peoples, and the legacy of injustices imposed on them and legitimated by the state. Key among these are the abrogation of indigenous sovereignty, the dispossession of land, and the purposeful destruction of indigenous cultures. As with gender roles, the social structure of heteropatriarchy, and class inequality, settler colonialism is actively reproduced through everyday practices, distinctions, and ideas ([Veracini 2010](#)). There is no technological 'fix' for colonialism, dispossession of land, and loss of culture, but we can still consider how AI systems "reproduce colonial ontology and epistemology" ([Costanza-Chock 2020, p. 67](#)), and what these systems would look like if they were designed to support indigenous sovereignty and resurgence.

**Conclusion**

The broader lesson here is the importance of attending to the specifics of inequality and social structure. Settler colonialism is a distinctive social arrangement, but it is not a more complex phenomenon that Black-White inequality in the U.S., which has its own ongoing history of structured injustice ([Omi and Winant 2014](#); [Bonilla-Silva 2015](#)), and which is interrelated with settler colonial logic ([Lloyd and Wolfe 2016](#)). In short, fairness algorithms are not generalizable and social inequalities are deep and multiple. The concept of bias is limiting and should often jettisoned where more specific conceptualizations of inequality are available. Rather than being concerned over how socio-technical systems reproduce pre-existing biases, we can actually name what we want to avoid reproducing: identifying processes, structures, hierarchies and concepts that have already been articulated by critical AI scholars and those in the social sciences and humanities. Being specific also helps us to name desirable alternatives to reproducing injustice, and orients us to where our actions can have meaningful impact.

Conservative approaches to AI ethics seek to achieve equal treatment under existing institutions, but enabling radical change (such as alternatives to prison and policing) is a tougher problem. A first step is appreciating that better ML models and AI systems will not provide an answer to what are political rather than technical problems ([Green 2018](#); [Wong 2020](#)). Given that politics is fundamentally about power, we would do well to recognize how these systems currently work to intensify, maintain, and optimize existing forms of power. Critical scholarship provides the tools to understand our complicity in upholding unjust and oppressive structures; racism, sexism, ableism, exist as processes rather than individual traits, and are maintained by the daily 'unintentional' actions of well-meaning people. Given our social circumstances, one cannot simply opt out of patriarchy, colonialism, or capitalism, but understanding how these structures work and are sustained can inform ways to limit our support, including through refusal (Cifor et al. 2019) and the work we choose not to do. The more difficult task, of actually shifting power (Kalluri 2020) and producing alternatives to existing injustices, may involve technical innovation – some kind of "radical data science" (Keyes 2019) with new 'use cases' or different ontologies. To this end, Sasha Costanza-Chock's (2020) guide for practitioners of "design justice" provides some valuable insights and connections. These approaches can help technologists work in a more responsible, socially-informed manner, but they are also grounded in the understanding that social issues are not technical problems amenable to technical fixes.

The latest manifestation of the Black Lives Matter movement led to unprecedented pressures for structural change in policing – enabled by social media hashtags and video sharing, but mobilized through fairly conventional forms of political action. While the structural changes that resulted have been limited to specific municipalities and police departments, this political movement has helped to broaden discussions beyond the problem of racial bias and towards a deeper understanding of systemic issues, opening room to articulate alternatives to existing structures (D'Souza 2020; Pettit 2020). Important work is being done to push AI development in a similar direction, but much more is needed to disrupt conservative complacency. This will require engaging with the foundational work of critical AI scholars, recognizing the inherently political qualities of new technologies, and drawing on forms of expertise (including political expertise) that have traditionally been out of scope in AI.

Crawford K, Joler V (2018) Anatomy of an AI System. In: Anat. AI Syst. http://www.anatomyof.ai. Accessed 7 Sep 2018

Cunneen C (2014) Colonial processes, Indigenous peoples, and criminal justice systems. In: Bucerius SM, Tonry MH (eds) The Oxford handbook of ethnicity, crime, and immigration. Oxford University Press, Oxford, pp 386–407

Davis G, Evans MJ (2018) Surgically Shaping Sex: A Gender Structure Analysis of the Violation of Intersex People's Human Rights. In: Risman BJ, Froyum CM, Scarborough WJ (eds) Handbook of the Sociology of Gender. Springer, Cham, pp 273–284

Doyle-Burke D, Smith J (2020) The Radical AI Podcast. https://www.radicalai.org. Accessed 7 Jul 2020

D'Souza S (2020) Activists in the U.S. claim partial victory in long battle to reform, defund police departments. In: CBC News. https://www.cbc.ca/news/world/police-defund-u-s-nypd-1.5637129. Accessed 10 Jul 2020

Edwards J (2006) The Radical Attitude and Modern Political Theory. Springer

Eubanks V (2017) Automating Inequality: How High-Tech Tools Profile, Police, and Punish the Poor. St. Martin's Press, New York, N.Y

Ferreira FHG, Peragine V (2015) Equality of Opportunity: Theory and Evidence. The World Bank

Friedler SA, Scheidegger C, Venkatasubramanian S (2016) On the (im)possibility of fairness. ArXiv160907236 Cs Stat

Friedman B, Nissenbaum H (1996) Bias in Computer Systems. ACM Trans Inf Syst 14:330–347. https://doi.org/10.1145/230538.230561

G7 Science Academies (2019) Artificial intelligence and society. https://rsc-src.ca/sites/default/files/Artificial%20intelligence%20and%20society%20G7%202019.pdf. Accessed 31 May 2019

Gandy OH (2009) Coming to Terms with Chance: Engaging Rational Discrimination and Cumulative Disadvantage. Routledge, London

Gillespie T, Seaver N (2016) Critical Algorithm Studies: a Reading List. In: Soc. Media Collect. http://socialmediacollective.org/reading-lists/critical-algorithm-studies/. Accessed 7 Jul 2020

Goertz G (2006) Social Science Concepts: A User's Guide. Princeton University Press

Green B (2019) "Good" isn't good enough. In: Proceedings of the AI for Social Good workshop at NeurIPS. Vancouver

Green B (2018) Data science as political action: grounding data science in a politics of justice. ArXiv Prepr ArXiv181103435

Green B, Hu L (2018) The Myth in the Methodology: Towards a Recontextualization of Fairness in Machine Learning. Stockholm